\documentclass[showkeys]{revtex4}

\usepackage{amsmath}
\usepackage{amsthm}
\usepackage{amsfonts}
\usepackage{graphicx}
\usepackage{multirow}

\newcommand{\degree}[0]{^{\circ}}

\newcommand{\Dir}[0]{\mathrm{Dir}}
\newcommand{\DP}[0]{\mathrm{DP}}

\newcommand{\CDP}[0]{\mathrm{CDP}}

\theoremstyle{plain}
\newtheorem{theorem}{\bf Theorem}[section]

\newtheorem{corollary}{\bf Corollary}[section]
\newtheorem{lemma}{\bf Lemma}[section]

\theoremstyle{definition}
\newtheorem*{definition}{Definition}
\newtheorem*{remark}{Remark}

\begin{document}
\title{Robust estimation of risks from small samples}

\author{
Simon H. Tindemans}
\email{s.tindemans@imperial.ac.uk}

\author{Goran Strbac}
\affiliation{Department of Electrical and Electronic Engineering, Imperial College London, 
London SW7 2AZ, UK}


\keywords{Rare event analysis, Bayesian inference, nonparametric methods, Dirichlet process, imprecise probabilities, resampling methods}


\begin{abstract}
Data-driven risk analysis involves the inference of probability distributions from measured or simulated data. In the case of a highly reliable system, such as the electricity grid, the amount of relevant data is often exceedingly limited, but the impact of estimation errors may be very large. This paper presents a robust nonparametric Bayesian method to infer possible underlying distributions. The method obtains rigorous error bounds even for small samples taken from ill-behaved distributions. The approach taken has a natural interpretation in terms of the intervals between ordered observations, where allocation of probability mass across intervals is well-specified, but the location of that mass within each interval is unconstrained. This formulation gives rise to a straightforward computational resampling method: Bayesian Interval Sampling. In a comparison with common alternative approaches, it is shown to satisfy strict error bounds even for ill-behaved distributions.
\end{abstract}

\maketitle


\section{Motivation: risk analysis for grids}

Managing a critical infrastructure such as the electricity grid requires constant vigilance on the part of its operator. The system operator identifies threats, determines the risks associated with these threats and, where necessary, takes action to mitigate them. In the case of the electricity grid, these threats include sudden outages of lines and generators, misprediction of renewable generating capacity, and long-term investment shortages.

Quantitative risk assessments often amounts to data analysis. In cases where directly relevant data is available (e.g. historical wind variability), it can be used directly to estimate future performance (hindcasting). In other cases, risk analysis is based on models, for example in distribution network outage planning. Even so, for sufficiently complex models, risks are rarely computed directly. Instead, Monte Carlo simulations are used to generate `virtual' observations, which are analysed statistically.

A particular challenge is the management of high-impact low-probability events. By definition, such events are exceedingly rare, but misjudging their likelihood of occurrence can have significant consequences. Cascading outages that can lead to partial or full system blackouts are a prime example of such events. In recent years, their modelling and analysis has become a very active area of research. Models are typically evaluated using Monte Carlo simulations \cite{AlizadehMousavi2012, Henneaux2012}, although other innovative approaches are also being developed \cite{Rezaei2015}.

The statistical validity of simulation outcomes is often not addressed, and when it is, this is typically by means of the standard error with the implicit invocation of the central limit theorem (e.g. \cite{AlizadehMousavi2012}). One recent exception is \cite{Dobson2013}, which explicitly asked how many blackout occurrences were required to estimate event probability with sufficient accuracy. However, this analysis was limited to a single binary outcome (i.e. a Bernoulli random variable). 

This paper contributes to closing this methodological gap, by introducing a method for the robust analysis of rare event data, which is applicable to generic real-valued observations. A simple  example in Section~\ref{sec:examples} illustrates how the method can be applied to cascading outage simulations. Critically, the method satisfies strict accuracy requirements, which will become increasingly important as operators move away from deterministic security standards (e.g. `$N-1$') to probabilistic standards that are enforced using Monte Carlo-based tools.

\section{Introduction}

We consider the problem of inferring  properties of a random variable $X$ from independent observations $\{x_{1:n}\}$. For example, one may attempt to estimate its expectation $E[X]$ or distribution $F_X^*$. This problem can be found in idealised form in the analysis of Monte Carlo simulations, where independence of observations is often guaranteed.

A range of approaches exist for addressing this elementary inference problem, differing in their underlying assumptions and applicability \cite{Robert1999}. On one end of the scale there are simple point estimates, such as the sample mean or - when a distribution is concerned - the empirical distribution. When more accuracy is required it is common practice to report a confidence interval, usually by means of an implicit invocation of the central limit theorem. However, this approach is not suitable when sample sizes are small or the distribution of $X$ is sufficiently heavy-tailed or skewed. In such cases the bootstrap method \cite{Efron1979} is a well-established alternative, but it is limited to resampling of observed values, so there is no way to account for features of the distribution that have not been sampled. This shortcoming is shared by its Bayesian variant, the Bayesian bootstrap \cite{Rubin1981}. It is particularly restrictive when sample sizes are very small, or for the analysis of rare events where a small fraction of the probability mass has a disproportionate effect on the parameter of interest.

Ferguson's Dirichlet process \cite{Ferguson1973} may be used for inference beyond strictly observed values, but this requires the selection of a prior measure. Furthermore, the resulting inferred (random) distributions are ultimately discrete. A different approach is taken by the Nonparametric Predictive Inference (NPI) method \cite{Coolen2006}, which makes use of imprecise probabilities to circumvent these issues. However, its definition as an incremental predictive process makes it less suitable for the inference of the underlying time-invariant distribution $F_X^*$ or for efficient computational implementations.

In this paper we introduce a robust method for inference from independent real-valued observations with minimal assumptions. The method relies only on the observations and the specification of a bounding interval $I$ for the random variable $X$ (which may be taken as the entire real line $\mathbb{R}$). It results in  conservative posterior distributions for any quantities under consideration, but credible intervals for such quantities remain valid even for ill-behaved distributions. This is a desirable feature for applications where accuracy is paramount, such as risk assessment of critical systems. Specific contributions of this paper are as follows:
\begin{itemize}
\item The Dirichlet process is reformulated in a way that separates its expectation and random fluctuations (Section \ref{sec:UDP})
\item A non-parametric robust posterior distribution is defined, making use of imprecise probabilities in the form of probability boxes (Section \ref{sec:posterior}).
\item Interval estimates for monotonic functions $q$ of $F_X^*$ (such as mean, median, etc.) are formulated (Section \ref{sec:q}), accompanied by a straightforward resampling implementation  (Section \ref{sec:resampling})
\item Examples illustrate the robust nature of the proposed method, including an application to blackout risk estimation (Section \ref{sec:examples}).
\end{itemize}

\section{Preliminaries}

\paragraph{Notation}
The summary notation $v_{j:k}$ is used to indicate a sequence of values $v_i$ with consecutive indices $i=j,\ldots,k$. Capital symbols refer to random variables or non-decreasing functions on the real line (e.g. cumulative distribution functions); it will be clear which is intended from the context. Caligraphic letters (e.g. $\mathcal{F}$) indicate \emph{random} non-decreasing functions, i.e. realisations of distributions-of-distributions.

\subsection{Dirichlet distribution}
Let us consider the problem of determining an unknown discrete (categorical) probability distribution $\mathrm{Pr}^*$ on a set of $k$ disjoint events $A_{1:k}$, i.e. $\sum_{i=1}^k \mathrm{Pr}^*(A_i)=1$. To estimate $\mathrm{Pr}^*$ from independent realisations of the corresponding random variable we may use Bayesian inference with the Dirichlet distribution as a prior distribution. The Dirichlet distribution $\Dir[\alpha_{1:k}]$ is parametrised by $\alpha_{1:k}$ and its probability density is structured as
\begin{align} \label{eq:DirichletDistributionDefinition}
f(p_{1:k-1}; \alpha_{1:k}) \propto \prod_{i=1}^{k} p_i^{\alpha_i -1},
\end{align}
with the constraint $\forall i: p_i \ge 0$. Note that the density is formally only defined as a function of $p_{1:k-1}$, and the final component $p_k$ is computed as  $p_k = (1-\sum_{i=1}^{k-1} p_i)$. Nevertheless we include $p_k$ in Eq.~\eqref{eq:DirichletDistributionDefinition} to highlight the symmetry of the expression. Permissible values of $\{p_i\}_{i=1}^k$ are restricted to the unit $(k-1)$-simplex $\Delta^{k-1}$ that is defined as
\begin{equation}\label{eq:unitsimplexdefinition}
\Delta^{k-1} = \left\{ \vec{p} \in \mathbb{R}^{k} :  p_i \ge 0, \forall i; \sum_{i=1}^{k} p_i =1 \right\}.
\end{equation}

The values $p_i$ represent the probabilities associated with the events $A_i$. The Dirichlet distribution thus  represents a continuous distribution on the space of $k$-dimensional discrete probability distributions. Let us define the random variables $\{P_{1:k}\}$ as
\begin{equation}
 \{ P_{1:k} \} \equiv \{ P(A_1),\ldots,  P(A_k) \}  \sim \Dir[\alpha_{1:k}]. 
\end{equation}
The expectation of the event probabilities is given by
\begin{equation}
E[P_{1:k}] = \frac{1}{\sum_{i=1}^k \alpha_i}\alpha_{1:k}.\label{eq:DirExpectation}
\end{equation}

Because the Dirichlet distribution is a conjugate prior, the Bayesian posterior distribution is also a Dirichlet distribution. Specifically, let us consider a set of observation counts $n_{1:k}$ for each event, with $n_i \in \{0, 1, \ldots \} $ and a prior distribution $\Dir[\alpha_{i:k}]$. The posterior distribution is then given by $\Dir[\beta_{1:k}]$ with $\beta_i = \alpha_i + n_i$. This illustrates that the parameters $\alpha_i$ may be considered `pseudo-observations' on the same footing as the actual observations $n_i$. As a result, the posterior expectation of $P_{1:k}$ conditional on the observations $n_{1:k}$ is
\begin{equation}
E[P_{1:k}|n_{1:k}] = \frac{1}{\sum_{i=1}^k (\alpha_i + n_i)} \{\alpha_i + n_i \}_{i=1}^k.
\end{equation}

The probability associated with the union of two disjoint events $A_i$ and $A_j$ is given by the sum of their probabilities $p(A_i)$ and $p(A_j)$. In the case of a Dirichlet distribution, these probabilities are random variables $P_i$ and $P_j$.  The Dirichlet distribution has the property that the resulting $(k-1)$-dimensional distribution is obtained simply by summing the relevant parameters $\alpha_i$ and $\alpha_j$.
\begin{equation}
\{P_{1:i-1},\boldsymbol{P_i + P_j}, P_{i+1:j-1}, P_{j+1:k}\}  
\sim \Dir[\alpha_{1:i-1},\boldsymbol{\alpha_i + \alpha_j}, \alpha_{i+1:j-1}, \alpha_{j+1:k} ]  \label{eq:DirichletCombination}
\end{equation}
Repeated application of this process until only the events $A_i$ and its complement $A_i^c$ are left (e.g. `heads' and `tails') yields 
\begin{equation}
\{P(A_i), P(A_i^c)\}  \sim \Dir[\alpha_i, \sum_{j \neq i} \alpha_j ].\label{eq:twoParamDirichlet}
\end{equation}
Because $P(A_i^c) = 1-P(A_i)$  by definition, we may simply consider the marginal distribution of the random variable $P(A_i)$, which is the beta distribution with parameters $\alpha_i$ and $\sum_{j \neq i} \alpha_j $
\begin{equation}
P(A_i) \sim  \beta[\alpha_i, \sum_{j \neq i} \alpha_j ]. \label{eq:DirichletBeta}
\end{equation}

\subsection{Dirichlet process}
Ferguson's Dirichlet process \cite{Ferguson1973} provides a natural extension of the discrete Dirichlet distribution to an infinite set of events, and the continuous domain $\mathbb{R}$ in particular. The Dirichlet process $\DP[m]$ on $\mathbb{R}$ is a distribution of probability distributions (i.e. a random probability measure) that is parametrised by the measure $m: \mathcal{B}(\mathbb{R}) \to \mathbb{R}^+$. It is defined by the condition
\begin{equation}
\mathcal{P} \sim \DP[m] \implies 
\{\mathcal{P}(A_1), \ldots, \mathcal{P}(A_n) \} \sim \Dir[m(A_1), \ldots, m(A_n)], \label{eq:DPdefinition}
\end{equation}
where $\mathcal{P}$ is a random probability measure and $\{ A_{1:n} \}$ is any arbitrary disjoint covering of $\mathbb{R}$. It deserves mention that even when the measure $m$ has a continuous density on $\mathbb{R}$, samples from the corresponding Dirichlet process distribution will have discrete features \cite{Ferguson1973} (see also Figure \ref{fig:IDP}).

Like the Dirichlet distribution for discrete distributions, the Dirichlet process on the real line is a conjugate prior for Bayesian inference \cite{Ferguson1973}. Using $\mathcal{P}_0 \sim  \DP[m_0]$ as a prior, and $n$ independent observations $ \{ x_{1:n}\}$, it follows from \eqref{eq:DPdefinition} that
\begin{equation}
\mathcal{P}_n \sim \DP\left[m_0 + \sum_{i=1}^n \delta_{x_i}\right].\label{eq:DPBayesian}
\end{equation}
Here, $\delta_{x_i}$ are measures corresponding to a unit mass located at $x_i$, within the support set of $m_0$.

\subsection{Probability box representation of imprecise probabilities}

In probability theory as defined using Kolmogorov's axioms, each event $A$ has a specific probability $p_A$. When our state of knowledge does not permit us to make such precise statements one can resort to imprecise probabilities \cite{Walley1991}: instead of the value $p_A$ we assign an interval probability consisting of a pair of lower and upper probability bounds $\underline{p}_A$ and $\overline{p}_A$.

A particular representation of imprecise probabilities on $\mathbb{R}$ is the  \emph{probability box}\cite{Ferson2003} (also summarised as p-box), defined by a lower probability distribution $\underline{F}(x)$ and an upper probability distribution $\overline{F}(x)$.  We write
\begin{equation}
F^{\diamond} \equiv [\underline{F} , \overline{F} ],
\end{equation}
as a shorthand for the set of distribution functions enclosed by $\underline{F}$ and $\overline{F}$: 
\begin{equation}
F^{\diamond}  = \{F : \forall x \in \mathbb{R}: \underline{F}(x) \le F(x) \le \overline{F}(x) ; \forall \varepsilon > 0: F(x+\varepsilon) \ge F(x) \} \label{eq:pboxbounds}
\end{equation}

Probability boxes define a set of permissible distribution functions, but do not assign any preference of probabilities to set members. A probability box associates a \emph{probability interval} with the event $X \in (a,b]$. The upper and lower bounds of this interval are given by:
\begin{subequations}  \label{eq:intervalProbBounds}
\begin{align}
\overline{P}_{(a,b]} & = \overline{F}(b) - \underline{F}(a)  \\
\underline{P}_{(a,b]} & = \mathrm{max}(\underline{F}(b) - \overline{F}(a), 0).
\end{align}
\end{subequations}

Probability boxes may also be interpreted in terms of possibility theory \cite{Flage2012} or belief functions in Dempster-Shafer theory \cite{Ferson2003}. Defined only by a pair of upper and lower distributions, the probability box is restricted to representing uncertainty in terms of simple intervals. Whereas every probability box can be expressed as a belief function, the converse is not true \cite{Ferson2003}. For example, it cannot be used to represent the knowledge that a random variable $X$ has a known `dead band' between $x_1$ and $x_2$. As such it is not the most general representation of imprecise probabilities, but it is sufficient for the purpose of our analysis.

\section{Inference using the Unit Dirichlet Process} \label{sec:UDP}

In this section we reformulate the Dirichlet process to disentangle its expected distribution and the random fluctuations around that distribution, and we use this formulation to restate the Bayesian posterior distribution for the Dirichlet process. 

\begin{definition}[Cumulative Dirichlet Process]For a given Dirichlet process  $\mathcal{P} \sim \DP[m]$ on the real line, the corresponding Cumulative Dirichlet Process (CDP) is parametrised by the cumulative function
\begin{equation}
M(x) =  m((-\infty,x]).
\end{equation}
$\CDP[M]$ is defined such that $\mathcal{F} \sim \CDP[M]$ satisfies the cumulative identity
\begin{equation}
\mathcal{F}(x) \stackrel{d}{=}  \mathcal{P}((-\infty, x]),
\end{equation}
where $\stackrel{d}{=}$ denotes equality in distribution. 
\end{definition}

\begin{lemma} Let $\mathcal{F} \sim \CDP[M]$, with $lim_{x\to\infty}M(x)<\infty$.  We express $M(x) = \alpha F(x)$ in terms of a scalar concentration parameter $\alpha = lim_{x\to\infty}M(x)$ and a cumulative probability distribution $F(x)$. Then the following properties hold:
\begin{align}
E[\mathcal{F}] =& F. \label{eq:CDPExpectation}\\
\lim_{\alpha \to \infty} \mathcal{F} = & F. \label{eq:CDPLimit}
\end{align}
\end{lemma}

\begin{proof} 
$\mathcal{F}(x)=\mathcal{P}((-\infty,x])$, which, following \eqref{eq:DirichletBeta}, is distributed according to a beta distribution $\beta[\alpha F(x),\alpha(1-F(x))](x)$. Therefore, $\forall x: E[\mathcal{F}(x)]=F(x)$ so that $E[\mathcal{F}]=F$. In addition, it follows from the properties of the beta distribution that $\lim_{\alpha \to \infty}\mathrm{Var}(\mathcal{F}(x))=0$, so that $\lim_{\alpha \to \infty} \mathcal{F} = F$.
\end{proof}

\begin{definition}[Unit Dirichlet Process]Let the \emph{Unit Dirichlet Process} be the CDP that is generated by the concentration parameter $\alpha$ and the identity function on the unit interval $[0,1]$:
\begin{equation}
\mathcal{U}_{\alpha} \sim  \CDP[\alpha\; \mathrm{Id}_{[0,1]}].\label{eq:unitDPdefinition}
\end{equation}
\end{definition}

\begin{figure}[h]
\centering
\includegraphics[width=5in]{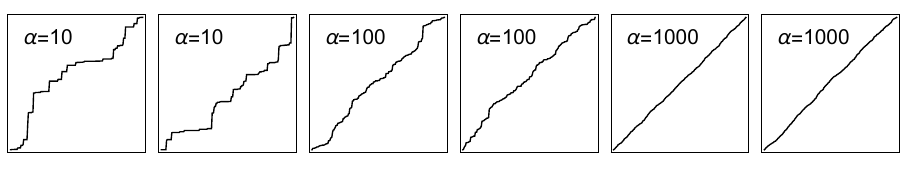}
\caption{Realisations of the identity Dirichlet process $\mathcal{U}_{\alpha}$. The value of $\alpha$ used to generate the realisation is shown in each panel. Two realisations are shown for each value of $\alpha$, illustrating its intrinsic variability. The plots were generated by discretising the interval $[0,1]$ into 200 intervals and using the discrete Dirichlet distribution $\Dir[\{\alpha/200\}_{i=1}^{200}]$ to sample the associated weights.} \label{fig:IDP}
\end{figure}

The unit Dirichlet process may be interpreted as a random non-decreasing map from the interval $[0,1]$ to $[0,1]$. From \eqref{eq:CDPExpectation} and \eqref{eq:CDPLimit} it follows that
$E[\mathcal{U}_{\alpha}]  =  \mathrm{Id}_{[0,1]}$ and $
\lim_{\alpha \to \infty} \mathcal{U}_{\alpha} = \mathrm{Id}_{[0,1]}$. Figure \ref{fig:IDP} shows several realisations of the random distribution $\mathcal{U}_{\alpha}$. The case $\alpha=10$ clearly demonstrates the discrete nature of the Dirichlet process, whereas $\alpha=1000$ illustrates the approximation of the identity map for $\alpha \to \infty$. Finally, The Unit Dirichlet Process has a tidy expression in terms of the `stick breaking process' definition of the Dirichlet process \cite{Sethuraman1994}:
\begin{equation}
\mathcal{U}_{\alpha} = \sum_{i=1}^{\infty} H_{U_i} B_i \prod_{j=1}^{i-1} (1-B_j).
\end{equation}
Here $H_c$ is the unit step function at $x=c$, and the sequences of independent random variables $\{U_{1:\infty}\}$ and $\{B_{1:\infty}\}$ are defined as $U_i \sim \mathrm{uniform[0,1]}$ and $B_i \sim \beta[1,\alpha]$.

\begin{theorem}\label{th:UDP}
Every Cumulative Dirichlet Process $\mathcal{F} \sim \CDP[\alpha F]$ can be expressed as
\begin{equation}
\mathcal{F} = \mathcal{U}_{\alpha} \circ F\label{eq:unitDPequivalence}
\end{equation}
where $\mathcal{U}_{\alpha}$ is a Unit Dirichlet Process and the $\circ$ symbol indicates composition of functions. 
\end{theorem}

\begin{proof}
Consider the cumulative Dirichlet process $\mathcal{F}\sim \CDP[\alpha F]$ and an arbitrary partitioning of the real line into half-open intervals $\{A_{1:n}\}$. This partioning is defined by the ordered set of points $\{y_{0:n}\}$, with $y_0=-\infty$, $y_{n}=+\infty$ and $A_i=(y_{i-1},y_i]$. Together, the set $\{y_{0:n}\}$ and $\mathcal{F}$ generate the non-decreasing sequence of random variables 
\begin{equation}
\{Y_{0:n}\} = \{\mathcal{F}(y_i)\}_{i=0}^n.  \label{eq:YFlink}
\end{equation}
The difference $Y_{i}-Y_{i-1}$ between two adjacent variables is the random probability $\mathcal{P}(A_i)$ associated with interval $A_i$. It follows from \eqref{eq:DPdefinition} that 
\begin{equation}
\{\mathcal{P}(A_i)\}_{i=1}^n \sim \Dir[\{ \alpha ( F(y_i) - F(y_{i-1})) \}_{i=1}^n].
\end{equation}
Defining $\{u_{0:n}\} \equiv \{ F(y_i)\}_{i=0}^n$, this can be written as
\begin{equation}
\{\mathcal{P}(A_i)\}_{i=1}^n \sim \Dir[\{ \alpha ( u_i - u_{i-1}) \}_{i=1}^n].
\end{equation}
Therefore, inverting the steps above,
\begin{equation}
\{Y_{0:n}\} = \{\mathcal{U}_{\alpha}(u_i)\}_{i=0}^n
\end{equation}
and a comparison with \eqref{eq:YFlink} gives
\begin{equation}
\{ \mathcal{F}(y_i) \}_{i=0}^n \stackrel{d}{=} \{ \mathcal{U}_{\alpha}(F(y_i)) \}_{i=0}^n.
\end{equation}
This holds for any partioning $\{ y_{0:n} \}$, so the equality can be extended to the entire domain. 
\end{proof}

Theorem \ref{th:UDP} neatly splits the cumulative Dirichlet process $\mathcal{F}$ into the expected shape ($F = E[\mathcal{F}]$) of the distribution and its inherent randomness ($\mathcal{U}_{\alpha}$). It shows that $\mathcal{F}$ can be understood as a random distortion of the generating distribution $F$ with fluctuations that decrease in amplitude as $\alpha$ increases. The same procedure can be applied to the Bayesian posterior of the Dirichlet process.

\begin{corollary}
Let $\mathcal{F}_0 \sim \CDP[\alpha F_0]$ be a prior CDP. Consider observations $\{x_{1:n}\}$ with an empirical distribution function 
$\hat{F}_n = \frac{1}{n} \sum_{i=1}^n H_{x_i}$,
where $H_c$ denotes the unit step function at $x=c$.
Restating \eqref{eq:DPBayesian}, the posterior CDP is given by
\begin{equation}
\mathcal{F}_n \sim \CDP\left[ \alpha F_0 + \sum_{i=1}^n H_{x_i} \right]. \label{eq:CDPBayesian}
\end{equation}
Invoking Theorem \ref{th:UDP}, this can be restated in the form
\begin{subequations} \label{eq:IDPBayesian} 
\begin{equation}
\mathcal{F}_n = \mathcal{U}_{\alpha + n} \circ F_n,
\end{equation}
where
\begin{equation}
F_n = \frac{\alpha}{\alpha + n} F_0 + \frac{n}{\alpha + n} \hat{F}_n.
\end{equation}
\end{subequations}
\end{corollary}

The implication of the decomposition \eqref{eq:IDPBayesian} is that the (cumulative) posterior distribution $\mathcal{F}_n$ may be considered a random deformation ($\mathcal{U}_{\alpha+n}$) of a weighted combination of the prior ($F_0$) and empirical ($\hat{F}_n$) distributions. 

\section{Robust inference of distributions} \label{sec:posterior}

The tools developed in the previous section can be applied to the central challenge in this paper: inferring the distribution $ F_X^*$ from a set of $n$ independent observations $\{  x_{1:n} \}$ of $X \sim F_X^*$. The prior assumption is that $X$ is restricted to an interval $I=[x_L,x_R]$. This lack of knowledge is represented by the \emph{vacuous} p-box
\begin{equation}
F^{\diamond}_0 = [H_{x_R},  H_{x_L} ]. \label{eq:vacuousprior}
\end{equation}
Recall that $H_c$ is the unit step function at $x=c$ and that $x_L$ and $x_R$ may take the values $-\infty$ and $\infty$, respectively.

\begin{definition}[P-box transformation] Let $\mathfrak{F}$ be the space of all distribution functions on the interval $I$, and $\mathfrak{F}^{\diamond}$ the space of probability boxes on $I$. Let $\mathfrak{U}$ be the set of all non-decreasing unit maps $U: [0,1] \rightarrow [0,1]$ with $U(0)=0$ and $U(1)=1$. 
For every $U \in \mathfrak{U}$ we define (with a slight abuse of notation) an identically named map $U: \mathfrak{F}^{\diamond} \rightarrow \mathfrak{F}^{\diamond}$ so that
\begin{equation}
U\circ F^{\diamond} \equiv [U\circ \underline{F}, U \circ \overline{F}],
\end{equation}
for any probability box $F^{\diamond} = [\underline{F},\overline{F}]$, with $\underline{F}, \overline{F} \in \mathfrak{F}$. The map $U \in \mathfrak{U}$ is order-preserving, so the transformed upper and lower bound distributions are sufficient to define the transformed p-box. In the following, the appropriate domain of the map $U$ ($[0,1]$ or $\mathfrak{F}^{\diamond}$) will be clear from the context.
\end{definition}

The Unit Dirichlet Process $\mathcal{U}_{\alpha} \sim \CDP[\alpha\; \mathrm{Id}_{[0,1]}]$ is a random element of $\mathfrak{U}$. It therefore defines a p-box randomisation in an analogous manner:
\begin{equation}
\mathcal{U}_{\alpha} \circ F^{\diamond} = [ \mathcal{U}_{\alpha} \circ \underline{F} ,\mathcal{U}_{\alpha} \circ \overline{F} ]  \label{eq:UDPtransform}
\end{equation}
Note that the UDP acts identically on the lower and upper bounds (each realisation transforms both identically). 

\subsection{Robust posterior distributions}
\begin{theorem} Define the $s$-robust posterior distribution as
\begin{align}
\mathcal{F}_n^{\diamond(s)} &= \mathcal{U}_{n+s} \circ F_n^{\diamond(s)}, \qquad \mathrm{with} \\
F_n^{\diamond(s)} &= \frac{s}{n+s}F_0^{\diamond} + \frac{n}{n+s}\hat{F}_n.
\end{align}
Then:
\begin{enumerate}
\item $\mathcal{F}_n^{\diamond(s)}$ is a posterior distribution corresponding to the vacuous prior $F_0^{\diamond}$ \eqref{eq:vacuousprior} with concentration parameter $s$ and observations $x_{1:n}$. 
\item $\mathcal{F}_n^{\diamond(s)}$ is a consistent estimator that converges to $F_X^*$ as $n \to \infty$. 
\end{enumerate}
\end{theorem}
\begin{proof}
(i) follows from \eqref{eq:IDPBayesian}, \eqref{eq:vacuousprior} and \eqref{eq:UDPtransform} and the fact that $\mathcal{U}_{\alpha}$ is order-preserving. Although the upper and lower bounds of the vacuous prior are strictly speaking degenerate as CDP priors, this can be regularised by using the limit $\varepsilon \downarrow 0$ of $\CDP[(1-\varepsilon)F_{\mathrm{degenerate}} + \varepsilon F^*_X]$.
(ii) Consistency follows from the fact that $\lim_{n\to\infty}F_n^{\diamond(s)}=\hat{F}_n=F_X^*$ and $\lim_{n\to\infty} \mathcal{U}_{n+s} = \mathrm{Id}_{[0,1]}$, thus $\lim_{n\to\infty} \mathcal{F}_n^{\diamond} = F_X^*$.
\end{proof}

\begin{remark}
In the limit $s \downarrow 0$ the upper and lower distributions coalesce into the single random distribution 
\begin{equation}
\mathcal{F}_{n}^{(s=0)} = \mathcal{U}_n \circ \hat{F}_n \sim \CDP\left[ n \hat{F}_n \right]. \label{eq:BBootstrap}
\end{equation}
This is the Bayesian Bootstrap method introduced by Rubin \cite{Rubin1981} as a Bayesian analog of the bootstrap method. It has assigns a (random) probability mass only to observed values $x_{1:n}$, so it has no ability to extrapolate beyond the lowest and highest observed values, and the posterior is always discrete.
\end{remark}

\begin{remark} 
Conceptually, the $s$-robust posterior distribution is closely related to the posterior distribution obtained using a Robust Bayes approach. However, it differs in the way it embeds continuous functions in the posterior. Using a prior set $F_0 \in F_0^{\diamond}$ and concentration parameter $s$, the Robust Bayes posterior can be written as $\mathcal{U}_{n+s}\circ F$, with $F \in F_n^{\diamond(s)}$. Although its range is identical to that of the $s$-robust posterior distribution, the discrete nature of the Dirichlet process means that members of the posterior distribution are smooth with zero probability. In contrast, the $s$-robust posterior distribution is a random p-box. Although its bounds have discrete features, continuous functions that fit within the p-box are equally part of the posterior distribution. 
\end{remark}

The parameter $s$ determines the relative weight of the prior and observations. It can therefore be considered the inverse of an effective `learning rate': small values result in faster convergence, but picking a value that is too small leads to an underappreciation of parts of the distribution that have not been observed, as represented by the vacuous prior.

\begin{theorem}\label{th:continuity} Let $\mathcal{F}_n^{\diamond(s)}$ be an $s$-robust posterior distribution. The posterior is compatible with continuous distributions $F_X^*$ only if $s \ge 1$. 
\end{theorem}
\begin{proof}
If the original distribution $F_X^*$ is (locally) continuous, the observations $\{x_{1:n}\}$ are almost surely distinct. Take the $i$-th smallest observation $x_{(i)}$ and consider the posterior distributions at $x_{(i)}\pm \varepsilon$. To avoid necessary discontinuities in the posterior distribution, the upper bound distribution at $x_{(i)} - \varepsilon$ must equal or exceed the lower bound distribution at $x_{(i)} +\varepsilon$:
\begin{equation}
\lim_{\varepsilon \downarrow 0} \mathcal{\overline{F}}_n^{(s)}(x_{(i)} - \varepsilon) \stackrel{?}{\succeq} \lim_{\varepsilon \downarrow 0} \mathcal{\underline{F}}_n^{(s)}(x_{(i)} + \varepsilon) \label{eq:localorderrequirement}
\end{equation}
Because the LHS and RHS distributions are linked through the order-preserving random map $\mathcal{U}_{n+s}$, the inequality should be interpreted as first order stochastic dominance \cite{Hadar1969} (ordering of quantiles). Making use of \eqref{eq:DirichletBeta}, \eqref{eq:localorderrequirement} can be expressed in terms of beta distributions
\begin{subequations} 
\begin{align} \label{eq:YZcondition}
Y & \stackrel{?}{\succeq} Z \\
Y \sim \beta[i - 1 + s, n - i + 1]   \qquad ; &  \qquad Z \sim \beta[i, n - i + s]  
\end{align}
\end{subequations}
It follows from the properties of the beta distribution that \eqref{eq:YZcondition} can only hold if $s \ge 1$.
\end{proof}

\begin{definition} The \emph{robust posterior distribution} $\mathcal{F}_n^{\diamond}$ is defined as the $s$-robust posterior distribution with $s=1$. 
\begin{align}
\mathcal{F}_n^{\diamond} &= \mathcal{U}_{n+1} \circ F_n^{\diamond}, \qquad \mathrm{with} \label{eq:Fdefinition} \\
F_n^{\diamond} &= \frac{1}{n+1}F_0^{\diamond} + \frac{n}{n+1}\hat{F}_n. \label{eq:posteriorInput}
\end{align}
\end{definition}

\begin{figure}[h]
\centering
\includegraphics[width=5in]{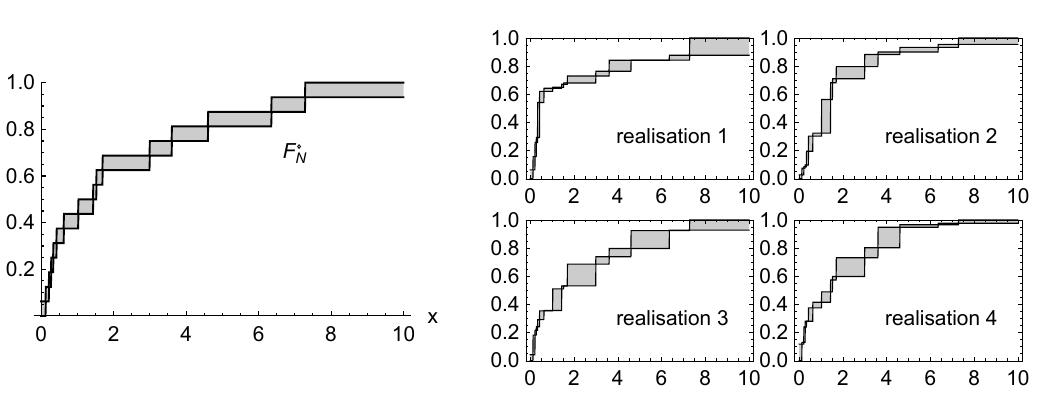}
\caption{The probability box $F_{n}^{\diamond}$ (left) has been constructed from the sample data in Table \ref{tab:smallData} and the interval $[0,\infty)$. Also shown (right) are four realisations of $\mathcal{F}_{n}^{\diamond}$ resulting from the random mapping of $F_{N}^{\diamond}$ by $\mathcal{U}_{n+1}$.}
\label{fig:combinedF}
\end{figure}

The `input' p-box $F^{\diamond}_{n}$ can be considered the spanning distribution of empirical distributions generated by the observations $\{x_{1:n}\}$ and one unknown observation in the interval $I$. Figure \ref{fig:combinedF} (left) depicts $F^{\diamond}_{n}$ for the observations in Table \ref{tab:smallData}. The right panel shows four random realisations of $\mathcal{F}_n^{\diamond}$. The realisations visually resemble a `chain of blocks' with the lower and upper bounds touching at each observation $x_{i}$. These upper and lower bound pairs force compatible distributions to traverse particular values at $x_{i}$ but leave them otherwise unconstrained, in line with theorem \ref{th:continuity}. The choice $s=1$ also ensures that an expected posterior probability of $1/(n+1)$ is assigned to each of the tail intervals $[x_L,x_{(1)}]$ and $[x_{(n)}, x_R]$. Further properties of $F^{\diamond}_{n}$, including comparisons with other methods, are discussed in the Supplementary Materials, Section S3.

\begin{table}[ht]
\centering
\caption{Sample data. These 15 observations are drawn independently from the log-normal distribution $\ln \mathcal{N}(0,1)$. For the robust analysis it is assumed that $I=[0,\infty)$.}
\begin{tabular}{l@{\hskip 0.5cm}l@{\hskip 0.5cm}l@{\hskip 0.5cm}l@{\hskip 0.5cm}l}
\hline 
1.435 & 0.276 & 3.603 & 0.211 & 2.996 \\ 
7.289 & 0.426 & 0.124 & 1.523 & 4.603 \\ 
1.696 & 0.620 & 0.338 & 6.351 & 1.026 \\
\hline
\end{tabular}
\label{tab:smallData}
\end{table}

\subsection{Interval-associated posterior probabilities}

The upper and lower bounds of the p-box $F_{n}^{\diamond}$ only take on $n+2$ distinct values. As a result, the (random) upper and lower values of the random probability box $\mathcal{U}_{n+1}\circ F_n^{\diamond}$ are given by $\mathcal{U}_{n+1}(v_i)$, with  $v_i = i/(n+1)$ and $i=0,\dots, n+1$. This partitioning allows for the projection of \eqref{eq:unitDPdefinition} onto an $n+1$-dimensional Dirichlet distribution. Making use of \eqref{eq:DPdefinition}, we can specify $\mathcal{F}_n^{\diamond}$ directly as a random p-box:
\begin{equation}
\mathcal{F}_{n}^{\diamond} = \left[\sum_{i=1}^{n+1} W_i H_{x_{(i)}} , \sum_{i=1}^{n+1} W_i H_{x_{(i-1)}}  \right] \label{eq:Fsimplified}
\end{equation} 
with $\{W_{1:n+1} \} \sim  \Dir \left[1,\ldots, 1\right]$ and $\{x_{(1):(n)} \}$ is the ordered data set $\{x_{1:n}\}$, augmented by $x_{(0)}=x_L$ and $x_{(n+1)}=x_R$.

The fact that the robust posterior distribution $\mathcal{F}_n^{\diamond}$ can be expressed in this simple manner has significant implications. Eq.~\eqref{eq:Fsimplified} provides an intuitive understanding of the method that has been developed. The interval $I$ is partitioned into $n+1$ closed intervals with boundaries at the observed points $\{x_i\}_{i=1}^N$. Each of these \emph{intervals} is assigned a random weight $W_i$, drawn from a Dirichlet distribution. Note that this Dirichlet distribution with $N+1$ unit parameters is in fact the uniform distribution on the unit simplex $\Delta^N$, as can be seen from Eqs.~\eqref{eq:DirichletDistributionDefinition} and \eqref{eq:unitsimplexdefinition}. 

Although a specific probability distribution governs the probability mass assigned to each interval, the  method provides no guidance regarding the way this mass is distributed \emph{within} each interval. This dichotomy can be interpreted as follows. The limited number of observations can only provide meaningful information about the large scale features (probability associated with intervals) of the probability distribution $F_X^*$. Without additional observations or assumptions no substantiated statements can be made about the features of $F_X^*$ on a smaller scale.

\section{Estimating population parameters} \label{sec:q}

In practical applications the object of interest is often not the distribution $F_X^*$, but a particular real-valued function $q^*=q[F^*_X]$ of that distribution, such as the expectation,  median or various measures of tail risk. In the context of this paper, we restrict ourselves to the case where $q$ is \emph{monotonic}. 

\begin{definition} A function $q: \mathfrak{F} \to \mathbb{R}$ is monotonic if the following is true for any two random variables $Y \sim F_Y$ and $Z \sim F_Z$:
\begin{equation}
Y \succeq Z \Rightarrow q[F_Y] \ge q[F_Z].
\end{equation} 
Where the partial ordering '$\succeq$' denotes first order stochastic dominance \cite{Hadar1969}, which is equivalent to the statement $\forall x \in \mathbb{R}: F_Y(x) \le F_Z(x)$, with a strict inequality holding for at least one point. 
\end{definition}

The strong condition of monotonicity holds for many basic population parameters, including the mean, quantiles (value-at-risk), and truncated means (conditional value-at-risk). More generally, it holds for the class of coherent risk measures \cite{Artzner1999}, although a customary minus sign is usually incorporated into the definition of monotonicity.

\begin{figure}[h]
\centering
\includegraphics[width=5.5in]{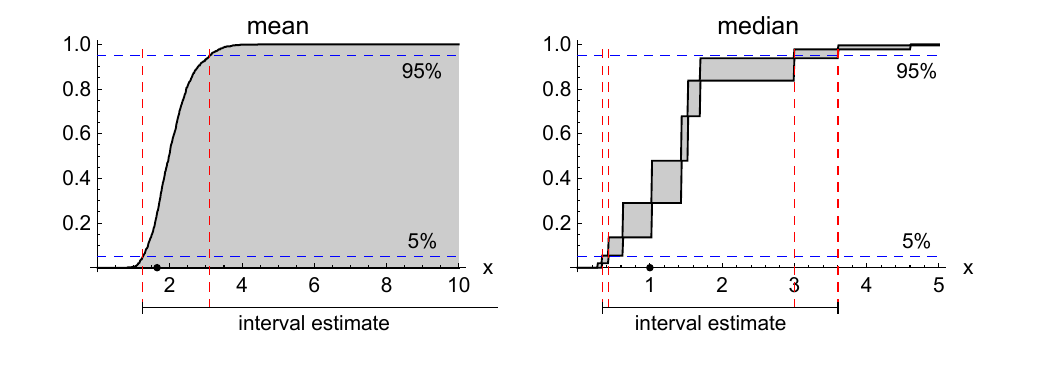}
\caption{Probability boxes for the mean $\mu$ and median $m$ of $X$, determined using the  sample data in Table \ref{tab:smallData}, the bounding interval $I=[0,\infty)$ and $n_{\mathrm{resample}}=1000$. The construction of the 90\% credibility interval ($c=0.9$) is illustrated as follows: dashed blue lines (horizontal) indicate the 5\% and 95\% quantiles, and dashed red lines (vertical) the corresponding values of $\mu$ and $m$. In case of the median (right), the two sets of vertical lines correspond to values generated using the upper and lower bounds of the random p-box $\mathcal{F}_n^{\diamond}$. The black dot indicates the true mean and median of the originating distribution. }
\label{fig:median}
\end{figure}

Consider the set of permitted $q$-values for a given p-box $F^{\diamond} = [\underline{F},\overline{F}]$, which we summarise by its minimum and maximum values. Due to the monotonicity of $q$, we have $
[q^{\mathrm{min}}, q^{\mathrm{max}}] = \left[ q[\overline{F}], q[\underline{F}] \right]$. For the \emph{random} p-box defined by $F^{\diamond}$, this induces the random $q$-interval
\begin{equation}
[Q^{\mathrm{min}}, Q^{\mathrm{max}}] = \left[ q[\overline{\mathcal{F}}_n], q[\underline{\mathcal{F}}_n \right]. \label{eq:Qrandomvars}
\end{equation}
The distributions of the bounding random variables $Q^{\mathrm{min}}$ and $Q^{\mathrm{max}}$ can be interpreted as the definition of a probability box for the population parameter $q^*$ (see also Figure~\ref{fig:median}):
\begin{equation}
F^{\diamond}_Q = [F_{Q^{\mathrm{max}}}, F_{Q^{\mathrm{min}}}] = [F_{q[\underline{\mathcal{F}}_n]}, F_{q[\overline{\mathcal{F}}_n]}]. \label{eq:Qdistribution}
\end{equation}

\paragraph{Reporting intervals}
In applications, it is usually preferable to summarise the estimate of a population parameter $q^*=q[F^*_x]$ in the form $[q_{\textrm{min}}, q_{\textrm{max}}]_c$, for a given credibility level $c \le 1$ (e.g. 95\%):
\begin{equation}
\mathrm{Pr}( q^* \in [q_{\textrm{min}},q_{\textrm{max}}]_c) \ge c \label{eq:qIntervalRequirement}
\end{equation}
In keeping with our robust approach, this is achieved by taking the \emph{span} of the $c$-credible values for the upper and lower bound processes, resulting in
\begin{equation}
[q_{\textrm{min}},q_{\textrm{max}}]_c = \left[ {F_{Q^{\mathrm{min}}}}^{-1}\left(\frac{1-c}{2}\right)  ,  {F_{Q^{\mathrm{max}}}}^{-1}\left(\frac{1+c}{2}\right) \right]. \label{eq:intervalEstimate}
\end{equation}
The construction of the interval estimate is illustrated in Figure \ref{fig:median}. Under the stated assumptions (independence of observations and the range of $X$ being confined to $I$) the credibility $c$ is a \emph{lower bound} for the probability that the real value is contained in the interval estimate. To emphasise both the accuracy and the inherent conservativeness of the interval estimate, this can be expressed as a "$c+$ credible interval" (e.g. 95\%+ credible interval). 

\section{Bayesian Interval Sampling} \label{sec:resampling}

The posterior distribution $\mathcal{F}_{n}^{\diamond}$ as expressed in \eqref{eq:Fsimplified} associates random probabilities $W_{1:n+1} \sim \mathrm{Uniform}(\Delta^N)$ with the intervals $[x_{(i-1)}, x_{(i)}]$, for $i=1,\dots,n+1$. This formulation suggests a computationally straightforward resampling method that can be used to probe the upper and lower bounds of $\mathcal{F}_{n}^{\diamond}$, and therefore (through \eqref{eq:Qrandomvars}) the distributions of $Q^{\mathrm{min}}$ and $Q^{\mathrm{max}}$. The algorithm, \emph{Bayesian Interval Sampling} (BIS), is described below.
\begin{enumerate}
\item For $i=1$ to $n_{\mathrm{resample}}$
\begin{enumerate}
\item Sample an $(n+1)$-dimensional weight vector $\{w_{1:n+1}^{(i)}\} \in \Delta^n$ from $\Dir \left[1,\ldots, 1\right]$
\item Compute $q^{\mathrm{min}}_i = q\left[\frac{1}{n+1}\sum_{j=1}^{n+1} w_j^{(i)} H_{x_{(j-1)}}\right]$
\item Compute $q^{\mathrm{max}}_i = q\left[\frac{1}{n+1}\sum_{j=1}^{n+1} w_j^{(i)} H_{x_{(j)}}\right]$
\end{enumerate}
\item Compute $\hat{F}_{Q^{\mathrm{min}}}$ as the empirical CDF of $\{q^{\mathrm{min}}_{1:n_{\mathrm{resample}}}\}$
\item Compute $\hat{F}_{Q^{\mathrm{max}}}$ as the empirical CDF of $\{q^{\mathrm{max}}_{1:n_{\mathrm{resample}}}\}$
\item Compute the $c$-credible interval estimate $[q_{\textrm{min}},q_{\textrm{max}}]_c$ using \eqref{eq:intervalEstimate}
\end{enumerate}

As a rule of thumb we use $n_{\mathrm{resample}}=100/(1-c)$ to ensure that we have 100 observations in the tails of the empirical distributions for $Q^{\mathrm{min}}$ and $Q^{\mathrm{max}}$. This ensures reasonable statistical stability for the computation of the interval $[q_{\textrm{min}},q_{\textrm{max}}]_c$. A modified algorithm for data sets with many identical observations is given in the Supplementary Materials, Section S1.

\begin{remark}
Bayesian Interval Sampling also has a natural interpretation in terms of the Bayesian bootstrap method \cite{Rubin1981}, a Bayesian variation of the (frequentist) bootstrap method \cite{Efron1979}. The connection between the algorithm above and the Bayesian bootstrap method is readily apparent through comparison with \eqref{eq:BBootstrap} and \eqref{eq:Qdistribution}. The upper bound distribution $\hat{F}_{Q^{\mathrm{min}}}$ for $q$ is the Bayesian bootstrap posterior distribution for the dataset $x_{1:n}$ augmented with a `pseudo-observation' of the lower domain boundary $x_L$. Similarly, $\hat{F}_{Q^{\mathrm{min}}}$ is the Bayesian bootstrap posterior distribution for $x_{1:n}$ augmented with the upper domain boundary $x_R$. In Bayesian Interval Sampling, both bounds are sampled simultaneously, guaranteeing sample-wise coherence ($q^{\mathrm{min}}_i \le q^{\mathrm{max}}_i$).
\end{remark}

\section{Examples}\label{sec:examples}

Bayesian Interval Sampling was used to analyse the data in Table \ref{tab:smallData}, using the bounds $I=[0,\infty)$. The population parameters considered were the median and the mean, resulting in the probability boxes in Figure~\ref{fig:median}. The probability box for the median (right) is well-behaved and results in a finite interval estimate. In contrast, the equivalent analysis for the mean (left) does not result in a finite interval estimate. This is caused by the lower bound distribution for the mean, which has a constant value of $F_{\mu^{\mathrm{max}}}=0$. This perhaps surprising result is a consistent outcome of our robust approach. No  assumption was made on the long tail behaviour of $F^*_X(x)$, other than a data-driven bound on the probability $Pr(X > x_{(n)})$. Therefore one cannot rule out the existence of a small probability mass at $x \to +\infty$, which makes an arbitrarily large contribution to the mean. This result also reflects the common understanding that the mean is not a robust statistic, whereas the median is. Further examples for the Value at Risk (VaR) and truncated mean are included in the Supplementary Materials, Section S2.

To illustrate the robustness of the Bayesian Interval Sampling method, its performance was analysed for the estimation of the mean in two variants of the (0,1)-log-normal distribution. The first variant truncated the distribution to the interval [0,50]. In realistic cases, such a  truncation could reflect the maximum size of $X$ imposed by a finite system size, for example the highest possible cost of a system malfunction. The second variant took the truncated distribution and additionally included a 1\% probability to observe the maximum value $x=50$ (equal to the upper bound of the truncated distribution). The inference performance on these distributions was assessed using 10,000 repeated experiments. For each experiment, 50 random observations were generated and used to compute an interval estimate for the mean. This interval was compared with the true mean of the distribution ($1.65$ and $2.13$, respectively), and a tally was kept to determine the overall accuracy of the method (i.e. proportion of correct predictions). Note that this is essentially a frequentist test, so both (frequentist) \emph{confidence} and (Bayesian) \emph{credible} intervals should satisfy \eqref{eq:qIntervalRequirement}. The accuracy of BIS was compared with the common Student $t$ and bootstrap approaches

\begin{table}
\centering
\caption{Comparison of methods for truncated log-normal distribution, with and without additional extreme events. Methods were applied as follows: (i) Student $t$ distribution: 95\% confidence interval based on the sample average and variance using Student's $t$ distribution with $n_{\mathrm{sample}}-1$ degrees of freedom. (ii) Bootstrap method: 95\% confidence interval based on bootstrap resampling with $n_{\mathrm{resample}}=2000$. (iii)  Bayesian Interval Sampling (BIS): 95\% credible interval based on  $n_{\mathrm{resample}}=2000$ and the bounding interval $[0,50]$.}
\label{tab:comparison}
\begin{tabular}{p{3.5cm}l|ccc}
distribution &method  & accuracy & median $q_{\mathrm{min}}$ & median $q_{\mathrm{max}}$ \\
& &  (target: 95\%) &  &  \\ 
\hline
\multirow{3}{\hsize}{Truncated log-normal \newline (mean: $1.65$)}& Student t & $90.3\%$  & 1.08 & 2.12\\
& Bootstrap &   $90.1\%$ & 1.15 & 2.14 \\
& BIS &  $98.7\%$ & 1.17 &  5.10 \\
\hline
\multirow{3}{\hsize}{Truncated log-normal \newline with extreme events \newline (mean: $2.13$)} & Student t &   $68.9\%$ & 0.98 & 2.60 \\
& Bootstrap &  $70.0\%$ & 1.20 & 2.64 \\
& BIS &  $98.8\%$ & 1.26 & 5.42  \\
\hline
\end{tabular}
\end{table}

The results are shown in Table \ref{tab:comparison}. Both the Student t and bootstrap methods have success rates well below 95\%, so neither method satisfies the objective stated in \eqref{eq:qIntervalRequirement}. This is a direct consequence of the long tail of the distribution, the impact of which is significantly exacerbated in the extreme event distribution. In contrast, the Bayesian interval sampling method has much lower misprediction rates of only 1.3\% and 1.2\%, well below the 5\% bound imposed by the 95\% credibility requirement. The rightmost columns of Table \ref{tab:comparison} show the median values of the lower and upper interval bounds (across 10,000 experiments). They illustrate that the improved accuracy of BIS is a direct consequence of more conservative interval estimates.

\paragraph{Estimation of blackout risks}

Finally, BIS was applied to the motivating example of estimating blackout risks in power systems. A simple minimal electrical network model was constructed as follows: a random $1000$-node topology was generated using a modified Barab\'asi-Albert model with non-preferential random attachment of two edges (transmission lines) for each new node, resulting in 1998 lines. Customer loads were uniformly allocated to nodes (with a value `1' in rescaled units), and an equal amount of generating capacity was distributed across the nodes using a Dirichlet distribution $\Dir[1/9,\ldots,1/9]$. The power flows across the transmission lines were computed using the DC power flow approximation, assuming identical reactances for all lines. Line capacities were initialised according to the common `N-1' security criterion, enforcing that single line failures do not lead to overloads. An additional 10\% margin was added to generator capacities and transmission lines.

For this random network, we investigated the risk (expected impact) of line outages that occur independently with a probability $\mathrm{P}_{\mathrm{outage}}=10^{-4}$ (in a given period of interest). The simultaneous occurrence of $k>1$ line outages can result in overloads in other lines: overloaded lines are tripped one by one until no more overloads are present. Simultaneously overloaded lines are tripped one by one in random order. If the network breaks up into multiple components, supply and demand are balanced in each island individually. If insufficient local generating capacity is available, load is shed uniformly across the nodes in the island. The total fraction of load shed is used as an indicator for the severity of the cascading event.

\begin{table}
\centering
\caption{Computation of loss-of-load risk. Stratified sampling was used for $k=2,\ldots,5$, on the basis of 10,000 $k$-line outages each. Robust 99\%+ estimates of the mean loss were computed using the Bayesian interval sampling method with $n_{\mathrm{resample}}=10,000$.}
\label{tab:blackouts}
\begin{tabular}{c|cl|l}
$k$ & $\mathrm{Pr}(k)$ & mean fraction of load lost & risk contribution \\
& & (99\%+ interval estimate)  &  ($\mathrm{Pr}(k) \times$ loss of load) \\
\hline
0, 1 & 0.98 & 0 (by design) & 0  \\
2 & $1.6\times 10^{-2}$ & [$1.5\times10^{-5}$, $6.1\times10^{-4}$] & [$2.5\times10^{-7}$, $9.9\times10^{-6}$]  \\
3 & $1.1\times 10^{-3}$ & [$4.7\times10^{-5}$, $5.9\times10^{-4}$]&    [$5.1\times10^{-8}$, $6.4\times10^{-7}$]\\
4 & $5.4\times 10^{-5}$ & [$2.0\times10^{-4}$, $1.0\times10^{-3}$]&    [$1.1\times10^{-8}$, $5.6\times10^{-8}$]\\
5 & $2.2\times 10^{-6}$ & [$3.0\times10^{-4}$, $1.3\times10^{-3}$]&    [$6.5\times10^{-10}$, $2.7\times10^{-9}$]\\
>5  & $7.4 \times 10^{-8}$ & [0,1] (no assumptions) & [0, $7.4 \times 10^{-8}$]   \\
\hline
Total & 1 & & [$3.1 \times 10^{-7}$,$1.1\times10^{-5}$] \\
\hline
\end{tabular}
\end{table}

The analysis was structured by conditioning on the number of simultaneous outages $k$. By design, $k=0$ and $k=1$ have no impact, and $k>5$ is treated as a potential full blackout (load loss can take any value in $[0,1]$), without further analysis. For intermediate $k$, the mean loss of load was estimated using 10,000 simulated $k$-line outages. 99\% credible intervals were computed using BIS. In this case, the The intermediate results and the computation of the overall 99\%+ risk estimate ( $[3.1\times10^{-7}, 1.1\times10^{-5}]$) are shown in Table~\ref{tab:blackouts}. 

\section{Discussion}

We have described a robust nonparametric Bayesian method to estimate properties of a distribution function $F^*_X(x)$ from a set of independent observations $\{ x_{1:n} \}$. It assumes only the existence of a bounding interval $I$, which may be taken equal to the real line. 

The resulting posterior distribution $\mathcal{F}_{n}^{\diamond}$ \eqref{eq:Fsimplified} is expressed as a random probability box. It has the intriguing feature that posterior probabilities are associated with the \emph{intervals} between observations, but the allocation of probability mass \emph{within} each interval is left undetermined. Therefore, one can consider $\mathcal{F}_{n}^{\diamond}$ as a probabilistic estimate of only the low-frequency (coarse grained) content of $F^*_X(x)$. As expected, the resolution of this low-frequency estimate improves with the sample size $n$. 

In applications, one often aims to infer real-valued properties $q$ of $F^*_X(x)$ (e.g. mean, median, any  coherent risk measure). We have shown how the posterior distribution $\mathcal{F}_{n}^{\diamond}$ induces a probability box for the estimation of $q*=q[F^*_X]$. In turn, this probability box can be used to define a robust credible interval for $q^*$. Due to the robust construction, this interval may be interpreted as the range of values for $q$ that is \emph{not ruled out} by the observations. The Bayesian Interval Sampling (BIS) method described in Section~\ref{sec:resampling} provides a straightforward resampling algorithm to perform this estimation. 

The imprecise nonparametric Bayesian method (and the BIS implementation) is uniquely suitable for applications that require strict accuracy bounds, where (i) the set of (relevant) data points is small (roughly 100 or less) or (ii) unobserved events can have a disproportionate effect on the quantity that is being estimated.

When only a small number of data points is available, the set of observations $\{ x_{1:n} \}$ may be significantly skewed compared to the generating distribution $F^*_X(x)$. The nonparametric Bayesian approach results in an inferred distribution for $F^*_X(x)$ that correctly accounts for such small sample variations. It should be noted that this same argument also applies when the total number of samples is large, but the number of \emph{relevant} samples is small. For example, this occurs when one characterises rare failures in reliable systems. 

Furthermore, the interval probability approach provides a robust framework to deal with unobserved events. Such events may be so-called high-impact low-probability events, which have a disproportionate impact on the population parameter $q^*$. In applications where it is necessary to analyse largely unobserved tails of a distribution it is generally advisable to use Extreme Value Theory (EVT) (see e.g. Coles \cite{Coles2001}). However, the application of EVT assumes that the long-tail behaviour of the process is sufficiently well-described by the data set. This may not be the case if the data set is very small, or when the underlying process is prone to exhibit very rare events that are qualitatively different from  other large events. Events of the latter type are sometimes referred to as `Dragon kings' \cite{Sornette2012}.

The method described in this paper is conservative by design, and the resulting uncertainty bands may be dishearteningly large (see e.g. Figure \ref{fig:median}). Conventional methods will usually provide estimates with tighter bounds, and their use may well be justified if the underlying distribution $F^*_X(x)$ is well-behaved. However, the robust posterior distribution (and BIS implementation) can still be used for comparison purposes. Significant differences in results will highlight the impact of explicit and implicit assumptions made in the analysis, pointing out the need to confirm their validity in a particular application.

\enlargethispage{20pt}

\section*{Acknowledgments}
The authors thank Martin Clark, Richard Vinter, Matthias Troffaes, Frank Coolen, Alastair Young and Chris Dent for thought-provoking discussions and helpful comments at various stages during the development of this research. The research leading to these results has received funding from the European Union Seventh Framework Programme (FP7/2007-2013) under grant agreement n$\degree$~274387.


\end{document}